\documentclass[conference,10pt]{IEEEtran}
\IEEEoverridecommandlockouts

\usepackage{cite}
\usepackage{amsmath,amssymb,amsfonts}
\usepackage{algorithmic}
\usepackage{graphicx}
\usepackage{textcomp}
\usepackage{xcolor}
\usepackage{booktabs}

\def\BibTeX{{\rm B\kern-.05em{\sc i\kern-.025em b}\kern-.08em
    T\kern-.1667em\lower.7ex\hbox{E}\kern-.125emX}}

\renewcommand{\baselinestretch}{1.05}
\begin{document}

\title{CLAP-Based Automatic Word Naming Recognition in Post-Stroke Aphasia\\
\thanks{This work was supported by the Swiss National Science Foundation
project CRSII5\_202228 on “Characterisation of motor speech disorders and
processes”}
}

\author{\IEEEauthorblockN{Yacouba Kaloga$^{1}$, Marina Laganaro$^2$, Ina Kodrasi$^{1}$}
\IEEEauthorblockA{
\textit{$^1$Signal Processing for Communication Group, Idiap Research Institute, Martigny, Switzerland} \\
\textit{$^2$Faculty of Psychology and Educational Science, University of Geneva, Geneva, Switzerland}
}
}

\maketitle

\begin{abstract}
Conventional automatic word-naming recognition systems struggle to recognize words from post-stroke patients with aphasia because of disfluencies and mispronunciations, limiting reliable automated assessment in this population. In this paper, we propose a Contrastive Language–Audio Pretraining (CLAP) based approach for automatic word-naming recognition to address this challenge by leveraging text–audio alignment. Our approach treats word-naming recognition as an audio–text matching problem, projecting speech signals and textual prompts into a shared embedding space to identify intended words even in challenging recordings. Evaluated on two speech datasets of French post-stroke patients with aphasia, our approach achieves up to 90\% accuracy, outperforming existing classification-based and automatic speech recognition-based baselines.
\end{abstract}


\section{Introduction}

Aphasia is an acquired language disorder that most often results from left hemisphere stroke~\cite{laska01_jim,azhar17_jspt}. Nearly one-third of stroke survivors are affected by aphasia~\cite{gronberg22_n}, experiencing impairments across multiple language domains such as speech production, auditory comprehension, reading, and writing~\cite{raymer15_oxford}. A particularly prominent and clinically relevant feature of all types of aphasia is anomia, which refers to word-finding difficulty and/or paraphasias~\cite{raymer15_oxford}.
Standardized clinical protocols such as the Boston Naming Test~\cite{kaplan2001boston} and the Western Aphasia Battery~\cite{kertesz2007western} assess anomia using confrontation naming tasks, where patients are shown pictures of objects and asked to name them. Confrontation naming tasks are also largely used in research protocols on aphasia assessment and
rehabilitation~\cite{Conroy_et_al_2018, Mason_Nickels_2022}.
While these assessments yield valuable clinical information, they are time-consuming and resource-intensive, requiring a clinician to listen to, transcribe, and code each response. Such a limitation has driven the development of automatic word naming recognition systems, which can automatically determine whether a patient has produced the target word and provide immediate feedback~\cite{ballard19_ajslp, tran22_rapid, rykova25_loquens, faliagka25_petra, abad12_interspeech, barbera20_interspeech}. Beyond supporting digital evaluation and therapy~\cite{palmer12_stroke}, such systems are also critical for voice-controlled and interactive assistive technologies, expanding accessibility for individuals with aphasia~\cite{salis16_apha}.

Existing automatic word naming recognition approaches can be broadly categorized into automatic speech recognition (ASR)-based approaches~\cite{ballard19_ajslp, tran22_rapid, rykova25_loquens, faliagka25_petra} and verification-based approaches~\cite{abad12_interspeech, barbera20_interspeech}. 
ASR-based approaches transcribe the speech signal and verify whether the expected word appears in the transcription. Many ASR-based approaches rely on off-the-shelf systems trained on neurotypical adult speech~\cite{ballard19_ajslp, rykova25_loquens, faliagka25_petra}, though fine-tuning on aphasic speech corpora has also been explored~\cite{tran22_rapid}. Despite their flexibility, ASR models often struggle with disfluent and paraphasic recordings. Even when trained on pathological corpora such as AphasiaBank~\cite{macwhinney11_aphasiology}, state-of-the-art ASR systems exhibit substantially higher word error rates (WER) for aphasic speech compared to neurotypical speech~\cite{tang23b_interspeech, sanguedolce24_interspeech}.
This lower performance is even more pronounced for word naming tasks, since isolated words without context provide no linguistic cues to aid recognition, challenging ASR models further.

Verification-based approaches, in contrast, do not require transcriptions and determine whether a spoken word matches a target  word using techniques such as acoustic likelihood matching~\cite{abad12_interspeech} or posteriorgram pattern matching~\cite{barbera20_interspeech}. While these methods have shown good performance in controlled settings, they are often ill-suited to clinical speech data. In recordings of productions from people with aphasia, the target word is often embedded within ongoing patient-clinician interaction and may be preceded or followed by multiple word naming attempts. This variability introduces ambiguity, making direct word-level matching unreliable.  
Moreover, verification-based approaches require retraining whenever a new target word is introduced, limiting their scalability for flexible therapy applications.

These limitations highlight the need for a new approach that can robustly identify the correctness of the word that the patient utters (even in the presence of overlapping speech, multiple naming attempts, or disfluencies) and that does not require retraining for each new target word. Inspired by the Contrastive Language–Audio Pretraining (CLAP) framework~\cite{elizalde23_clap}, we propose to reframe word naming recognition as a contrastive audio–text matching problem. In this formulation, each spoken attempt is embedded alongside descriptive textual labels (e.g., \emph{correct pronunciation of a target word} or \emph{mispronunciation}) in a shared representation space, allowing direct comparison between audio and text.
This contrastive setup provides several advantages over traditional ASR- and verification-based approaches. By using textual labels as anchors, the model focuses more on the intended word rather than the exact acoustic realization. As a result, it is inherently more robust to pronunciation variability, partial articulations, and interfering speech, since different realizations of the same word map close together in the embedding space whereas mispronunciations or unrelated speech map further away. Moreover, unlike verification-based approaches, a CLAP-based approach can generalize to unseen words without additional training, making it practical for real-world therapy scenarios.

For the implementation of the proposed CLAP-based word naming recognition system, we initialize the audio encoder with a pre-trained wav2vec2 model~\cite{w2v2} and the text encoder with DistilRoBERTa-base~\cite{Sanh2019}. Both encoders are then fine-tuned on aphasic speech datasets for the specific target words of interest using contrastive learning. Experimental results show that the proposed CLAP-based approach considerably outperforms ASR-based and classification baselines, demonstrating improved accuracy in identifying patients’ word naming accuracy despite pronunciation variability, disfluencies, and overlapping speech. 

\section{Dataset Description \& Problem Formulation}
\label{sec:dataset}

We use two anonymised French post-stroke aphasia speech datasets collected in the framework of two published research protocols~\cite{Demierre2025, Franco2025}, consisting of recordings from native French speaking adult patients suffering from post-stroke aphasia. Each recording captures a patient's
attempt to produce a target word in a confrontation naming
task presented on a computer screen via the E-Prime software (E-Studio), with audio recorded via a headset microphone.
The recordings are not pre-processed to remove potential clinician instructions, multiple naming attempts, or other potential recording artifacts.
An annotation by a speech and language pathologist indicates whether the produced word matches the target word.
All patients presented anomia in the framework
of mild to moderate aphasia as assessed with the French
assessment battery BECLA~\cite{Macoir2016}. 

The first dataset~\cite{Demierre2025} contains $7320$ recordings from $34$ patients ($14$ male and $20$ female).  
The mean age of the patients was $59.8$ years (range: $29$-$85$ years) and their average time from
stroke onset was $49$ months (range: $3$-$258$ months).
Each patient was prompted to produce $90$ unique target words (isolated nouns) multiple times. Among the $7320$ recordings, $6608$ are annotated as \emph{correct} by the clinician, indicating an overall accuracy of $90.3\%$ and suggesting that this patient group presents with relatively mild impairment and does not experience substantial difficulty with this task.

The second dataset~\cite{Franco2025} comprises $6608$ recordings from $16$ patients ($12$ male and $4$ female).  The mean age of the patients was $65.2$ years (range: $31$-$75$ years) and their average time from
stroke onset was $20$ months (range: $3$-$96$ months) Each patient was prompted to produce $56$ unique target words (isolated nouns) multiple times. There is minimal overlap between the target words in the two datasets (i.e., only 4), and hence, these datasets are separately analyzed in this paper. Among the $6608$ recordings, $3798$ are annotated as \emph{correct} by the clinician, indicating an overall accuracy of $57.5\%$ and suggesting that this patient group exhibits more moderate impairments and experiences noticeable difficulty with this task.

Our objective is to develop a robust system that classifies a spoken response as either a correct pronunciation of the prompted target word or a mispronunciation. To this end, we propose the CLAP-based approach outlined in Section~\ref{sec: clap} and compare it with two baseline approaches, i.e., a classification and ASR-based approaches described in Section~\ref{sec: baselines}.

\section{Proposed Approach}
\label{sec: clap}

We adopt a CLAP-style contrastive architecture that embeds audio and text into a shared representation space for word naming recognition. The audio modality is encoded using the wav2vec2 model~\cite{w2v2}. Frame-level audio representations are averaged across time and passed through a linear projection layer, resulting in a $d$-dimensional audio embedding $a_i \in \mathbb{R}^d$ for recording $i$. The text modality is encoded using DistilRoBERTa-base~\cite{Sanh2019}, using the representation of the first frame to represent the whole sequence as in~\cite{devlin2019}. This representation is projected to the same dimensionality $d$ via a linear layer, producing the final text embedding $t_i \in \mathbb{R}^d$.
Both audio and text embeddings are L2-normalized.

Text inputs are formulated as natural-language prompts describing the expected outcome based on the clinical annotation. Correctly pronounced responses receive a positive prompt that explicitly mentions the target word (e.g., "Correct pronunciation of the word \emph{pomme}"), while mispronounced responses receive a general negative prompt (i.e., "Mispronounced word").
This design enables the model to leverage linguistic structure to distinguish correct from incorrect naming attempts.

Formally, given a minibatch of $N$ paired audio--text training examples, we stack the audio embeddings into a matrix $A = [a_1,\dots,a_N]^\top \in \mathbb{R}^{N \times d}$ and the text embeddings into a matrix $T = [t_1,\dots,t_N]^\top \in \mathbb{R}^{N \times d}$. Training is performed using a symmetric contrastive objective based on cosine similarity. The similarity logits are defined as
\begin{equation}
L_{a\rightarrow t} = \tau_a A T^\top, \qquad
L_{t\rightarrow a} = \tau_t T A^\top,
\end{equation}
where $\tau_a=\exp(s_a)$ and $\tau_t=\exp(s_t)$ are learnable logit scales. Using identity labels $y_i=i$ (i.e., the $i$-th audio should match the $i$-th text), the model is optimized with the symmetric contrastive loss
\begin{equation}
\mathcal{L}=\frac{1}{2}\big[\mathrm{CE}(L_{a\rightarrow t}, y) + \mathrm{CE}(L_{t\rightarrow a}, y)\big],
\end{equation}
where $\mathrm{CE}$ denotes cross-entropy over in-batch negatives.  We evaluate this architecture with audio and text encoders initialized from pretrained models and fine-tuned using the contrastive objective on our  data~(cf. Section~\ref{sec: exp_setup}).

\section{Baseline Approaches}
\label{sec: baselines}
We compare the proposed CLAP-based approach against two baseline automatic word naming recognition approaches, i.e., a simple classification baseline and an ASR-based baseline.

\subsection{Classification}

For classification, each recording is assigned a discrete label. If the target word is pronounced correctly, the corresponding word is used as the label, and otherwise a generic \texttt{mispronounced} label is assigned. Speech signals are embedded using the wav2vec2 model~\cite{w2v2}. Frame-level representations are averaged across time to obtain a fixed-dimensional utterance embedding, which is then passed through a multilayer perceptron (MLP) to produce logits over the label set.
We evaluate this baseline using speech embeddings extracted from a pretrained wav2vec2 model. The embeddings are precomputed and fixed (i.e., the model weights are not fine-tuned), and only the MLP classifier is trained on our data (cf. Section~\ref{sec: exp_setup}).

\subsection{ASR}
For the ASR-based baseline, we perform automatic transcription using a connectionist temporal classification (CTC) head applied on top of wav2vec2 speech embeddings. A naming attempt is considered correct if the target word appears in the decoded transcription, and otherwise the response is classified as \texttt{mispronounced}. 
We fine-tune the wav2vec2+CTC model for ASR on our data (cf. Section~\ref{sec: exp_setup}). Following common practice for adaptation of wav2vec2 to ASR~\cite{w2v2, ot25}, the convolutional feature encoder is kept frozen, while the transformer encoder layers and the CTC projection layer are updated during fine-tuning.

\section{Experimental Settings}
\label{sec: exp_setup}

\subsection{Training}

Due to the limited number of speakers in the datasets, we employ a leave-one-speaker-out cross-validation strategy. In each fold, recordings from one speaker are held out as the test set, recordings from another randomly selected speaker are used as the validation set, and the remaining recordings form the training set. 
This procedure is repeated so that each speaker serves as the test speaker exactly once.
All models are trained for up to $30$ epochs with a batch size of $32$. Validation is performed every $5$ epochs, and the best model is selected based on validation performance using early stopping.  

{\emph{Classification baseline.}} For the classification baseline, we consider learning rates of $5 \times 10^{-5}$ and $1 \times 10^{-5}$. The optimal learning rate is selected based on the validation performance, which is assessed using the F1 score (cf. Section~\ref{sec: eval}). The model is implemented in PyTorch and optimized with the standard Adam optimizer~\cite{kingma2017adam}. 
The MLP consists of linear layer, batch normalization, \textsc{ReLU} activation, and a final linear layer for classification. 
The output dimension of the first linear layer is $256$, whereas the second linear layer’s output dimension equals the number of target classes.

{\emph{ASR-based baseline.}} \enspace For the ASR baseline, we use a learning rate of $5 \times 10^{-4}$. The model is implemented on HuggingFace and optimized with the AdamW optimizer~\cite{loshchilov2019}. Validation performance is assessed using WER. It should be noted that only utterances that are annotated as \emph{correct} are used for training ASR models.

{\emph{Proposed CLAP-based approach.}} For the proposed CLAP-based approach, we consider learning rates of $5 \times 10^{-5}$ and $1 \times 10^{-5}$ as for the classification baseline. The optimal learning rate is selected based on the validation performance, which is assessed using the F1 score (cf. Section~\ref{sec: eval}). The model is implemented in PyTorch and optimized with the standard Adam optimizer~\cite{kingma2017adam}.  The shared audio–text embedding space has a dimensionality of 256. The contrastive loss uses learnable logit scale parameters initialized with a temperature of 0.07, following standard practice in contrastive learning~\cite{wu2018un}. 

\subsection{Evaluation}
\label{sec: eval}
To evaluate the performance of the considered word naming recognition approaches, we report accuracy, precision, recall, and F1 score.
For our task, each possible label is treated as a separate class. The set of labels includes one for each target word (representing correct pronunciations) and an additional label representing mispronunciations. Accuracy is the proportion of recordings for which the model predicts the correct label. Precision indicates the proportion of predictions for a given class that are correct, recall indicates the proportion of true instances of a class that are correctly identified, and F1 score provides a balanced measure that considers both precision and recall.
Precision, recall, and F1 score are first computed per class and then averaged across all classes.  
These metrics provide a comprehensive assessment of model performance relative to the clinician-provided labels.

\subsection{Speech embeddings}

Speech embeddings from different layers of a pretrained self-supervised model such as wav2vec2 capture distinct features, making layer selection critical for downstream tasks~\cite{liang2025}. To determine the optimal speech representation for the word naming recognition task, we evaluated the performance of the classification baseline using embeddings from various layers and variants of wav2vec2. For this analysis, utterance-level representations were extracted from all 24 layers of several pretrained wav2vec2 models, including the large~\cite{baevski2020}, French~\cite{parcollet2024}, and multilingual (XLSR)~\cite{babu2021xlsr} variants. Layer-wise evaluation is restricted to the classification baseline due to computational complexity. While the ASR-based baseline and the proposed CLAP-based approach may benefit from different layers, a full exploration across all approaches is beyond the scope of this paper.

Our results indicated that using layer 12 of the French wav2vec2 model achieves the highest word naming recognition accuracy within the classification baseline (with layers 10–15 showing comparable performance). Consequently, for the results presented in Section~\ref{sec: exp_res}, we use embeddings extracted from layer 12 of the French wav2vec2  model for all considered approaches.

\section{Experimental Results}
\label{sec: exp_res}

\begin{table}[t!]
\centering
\caption{Mean and standard deviation of accuracy, precision, recall, and F1 score across all folds for the proposed CLAP-based approach and the baseline approaches on the first dataset.}
\label{tab: dataset1}
\setlength{\tabcolsep}{12pt}
\begin{tabular}{lccc}
\hline
Metric & Proposed & Classification & ASR \\
\hline
Accuracy   & \textbf{0.90 $\pm$ 0.07} & 0.85 $\pm$ 0.08 & 0.86 $\pm$ 0.04 \\
Precision & \textbf{0.90 $\pm$ 0.07} & 0.85 $\pm$ 0.09 & 0.85 $\pm$ 0.05 \\
Recall  & \textbf{0.93 $\pm$ 0.06} & 0.89 $\pm$ 0.09 & 0.90 $\pm$ 0.05 \\
F1 score   & \textbf{0.90 $\pm$ 0.07} & 0.85 $\pm$ 0.09 & 0.86 $\pm$ 0.05 \\
\hline
\end{tabular}
\end{table}

\begin{table}[t!]
\centering
\caption{Mean and standard deviation of accuracy, precision, recall, and F1 score across all folds for the proposed CLAP-based approach and the baseline approaches on the second dataset.}
\label{tab: dataset2}
\setlength{\tabcolsep}{12pt}
\begin{tabular}{lccc}
\hline
Metric & Proposed & Classification & ASR \\
\hline
Accuracy   & \textbf{0.85 $\pm$ 0.06} & 0.79 $\pm$ 0.08 & 0.80 $\pm$ 0.06 \\
Precision & \textbf{0.72 $\pm$ 0.21} & 0.65 $\pm$ 0.22 & 0.67 $\pm$ 0.16 \\
Recall  & \textbf{0.79 $\pm$ 0.23} & 0.69 $\pm$ 0.23 & 0.68 $\pm$ 0.19 \\
F1 score    & \textbf{0.74 $\pm$ 0.22} & 0.65 $\pm$ 0.23 & 0.66 $\pm$ 0.17 \\
\hline
\end{tabular}
\end{table}

Tables~\ref{tab: dataset1} and~\ref{tab: dataset2} summarize the performance of the proposed CLAP-based approach and of the baseline approaches for the first and second datasets, respectively.

On the first dataset, the CLAP-based approach clearly outperforms the baseline approaches, achieving a high accuracy of $0.90$ and a high F1 score of $0.90$. The recall of $0.93$ is particularly strong, showing that the model reliably identifies most correct and incorrect pronunciations. The precision of $0.90$ is slightly lower than the recall, indicating occasional misclassifications, but the overall balance between precision and recall as reflected by the F1 score remains robust. In comparison, the baseline approaches perform consistently worse across all metrics, highlighting the effectiveness of the CLAP-based approach in accurately distinguishing pronunciation correctness in this dataset.
It should be noted that the classification and ASR-based baselines show very similar performance, despite the ASR model being fully fine-tuned on aphasic speech. This confirms that even with fine-tuning on aphasic speech data, an ASR-based approach struggles to accurately transcribe pathological speech, performing similarly to training a simple classification layer on top of frozen embeddings.

On the second dataset, the CLAP-based approach continues to outperform the considered baselines, with even larger relative performance gains than on the first dataset across all metrics.
However, all approaches (i.e., the proposed approach and both baselines) perform worse on this dataset compared to the first dataset. We attribute this drop in performance to several factors related to dataset complexity, i.e., differences in the distribution and types of mispronunciations, in the amount of training data (with
the first dataset being larger than the second one, particularly with respect to the number of speakers), and in impairment severity. In particular, the second dataset includes more severely impaired patients, as indicated by the lower proportion of correct naming attempts (i.e., $57.5\%$ versus $90.3\%$ in the first dataset). 
Automatic methods may struggle in such conditions because more severely impaired speech often deviates more strongly from typical pronunciation patterns, exhibits more variability across patients, and contains ambiguous or partially correct utterances that models may misclassify. 
Investigating these factors further will be the focus of future work, as understanding them could guide targeted data collection and model design to improve generalization across diverse populations.

Overall, these results demonstrate that the proposed CLAP-based approach consistently provides the most reliable and effective performance across datasets, both in absolute terms and relative to baseline methods.

\section{Conclusion}
We proposed a CLAP-based approach for automatic word naming recognition in post-stroke aphasia, reframing the task as audio–text matching. This formulation avoids explicit transcription or pattern matching and improves robustness to disfluencies and overlapping speech, which are common challenges in aphasic recordings. Across two French aphasia datasets, our approach consistently outperformed classification- and ASR-based baselines, demonstrating its potential as a reliable tool for assessing language function in clinical and research settings. Future work will include a more detailed analysis of performance as a function of speaker characteristics and exploring novel training and prompting strategies to further improve recognition accuracy.

\bibliographystyle{IEEEtran}
\bibliography{refs}

@inproceedings{abad12_interspeech,
  title     = {Automatic word naming recognition for treatment and assessment of {A}phasia},
  author    = {Alberto Abad and Anna Pompili and Angela Costa and Isabel Trancoso},
  year      = {2012},
  booktitle = {Proc. Annual Conference of the International Speech Communication Association},
  pages     = {1055--1058},
  address = {Portland, USA},
  month = {Sept.}
}

@inproceedings{barbera20_interspeech,
  title     = {An utterance verification system for word naming therapy in {A}phasia},
  author    = {David S. Barbera and Mark Huckvale and Victoria Fleming and Emily Upton and Henry Coley-Fisher and Ian Shaw and William Latham and Alexander P. Leff and Jenny Crinion},
  year      = {2020},
  booktitle = {Proc. Annual Conference of the International Speech Communication Association},
  pages     = {706--710},
  month = {Oct.},
  address = {Shanghai, China}
}

@inproceedings{sanguedolce24_interspeech,
  title     = {When {W}hisper Listens to {A}phasia: {A}dvancing Robust Post-Stroke Speech Recognition},
  author    = {Giulia Sanguedolce and Sophie Brook and Dragos C. Gruia and Patrick A. Naylor and Fatemeh Geranmayeh},
  year      = {2024},
  booktitle = {Proc. Annual Conference of the International Speech Communication Association},
  pages     = {1995--1999},
  address = {Kos, Greece},
  month = {Sept.}
}

@inproceedings{tang23b_interspeech,
  title     = {A New Benchmark of {A}phasia Speech Recognition and Detection Based on {E-B}ranchformer and Multi-task Learning},
  author    = {Jiyang Tang and William Chen and Xuankai Chang and Shinji Watanabe and Brian MacWhinney},
  year      = {2023},
  booktitle = {Proc. Annual Conference of the International Speech Communication Association},
  pages     = {1528--1532},
  address = {Dublin, Ireland},
  month = {Aug.}
}

@article{macwhinney11_aphasiology,
  author  = {B. Macwhinney and D. Fromm and M. Forbes and A. Holland},
  title   = {{AphasiaBank: M}ethods for Studying Discourse},
  journal = {Aphasiology},
  year    = {2011},
  volume  = {25},
  number  = {11},
  pages   = {1286--1307},
  month = {Sept.}
}

@inproceedings{faliagka25_petra,
author = {Evanthia Faliagka and Manos Plitsis and Kosmas Palios and Nassos Katsamanis and Spyridoula Stamouli and Athanasia-Lida Dimou and Christos Panagiotou and Dimitris Karadimas and Christos Antonopoulos and Nikolas Voros and Vassilis Katsouros},
title = {Automatic assessment of verbal responses in a speech and language therapy platform},
year = {2025},
month = {June},
address = {Corfu Island, Greece},
booktitle = {Proc. ACM International Conference on PErvasive Technologies Related to Assistive Environments},
pages = {7–13},
}

@inproceedings{tran22_rapid,
author = {Trang Tran},
title = {Post-Stroke Speech Transcription Challenge ({T}ask {B}):
{C}orrectness Detection in {A}nomia Diagnosis with Imperfect Transcripts},
year = {2022},
month = {June},
booktitle = {Proc. {RaPID W}orkshop - Resources and ProcessIng of linguistic, para-linguistic and extra-linguistic Data from people with various forms of cognitive/psychiatric/developmental impairments},
address = {Marseille, France},
pages = {56--61},
}

@article{ballard19_ajslp,
  author  = {K. J. Ballard and N. M. Etter and S. Shen and P. Monroe and C. Tien Tan},
  title   = { Feasibility of Automatic Speech Recognition for Providing Feedback During Tablet-Based Treatment for {A}praxia of Speech Plus {A}phasia },
  journal = {American Journal of Speech-Language Pathology},
  year    = {2019},
  volume  = {28},
  number  = {2S},
  pages   = {818--834},
  month = {July}
}

@article{rykova25_loquens,
author = {Eugenia Rykova and Mathias Walther},
volume = {12},
year = {2025},
title = {Evaluation of German Automatic Speech Recognition solutions in the context of speech and language therapy support of people with {A}phasia},
journal = {Loquens},
}

@book{raymer15_oxford,
  title     = {The Oxford Handbook of Aphasia and Language Disorders},
  author    = {A. M. Raymer and L. J. Gonzalez Rothi},
  year      = 2015,
  publisher = {Oxford University Press},
  address   = {Oxford, United Kingdom},
 }

@book{kaplan2001boston,
  title        = {Boston Naming Test},
  author       = {E. Kaplan and H. Goodglass and S. Weintraub},
  year         = {2001},
  edition      = {2nd},
  publisher    = {Lippincott Williams \& Wilkins},
  address      = {Philadelphia, USA}
}

@book{kertesz2007western,
  author    = {A. Kertesz},
  title     = {Western Aphasia Battery -- Revised},
  year      = {2007},
  publisher = {Grune \& Stratton},
  address   = {New York, USA}
}

@article{azhar17_jspt,
  author    = {A. Azhar and S. Maqbool and G. A. Butt and S. Iftikhar and G. Iftikhar},
  title     = {Frequency of aphasia and its symptoms in stroke patients},
  journal   = {Journal of Speech Pathology \& Therapy},
  year      = {2017},
  volume    = {2},
  number    = {1},
}

@article{laska01_jim,
  author    = {A. C. Laska and A. Hellblom and V. Murray and T. Kahan and M. Von Arbin},
  title     = {Aphasia in acute stroke and relation to outcome},
  journal   = {Journal of Internal Medicine},
  year      = {2001},
  month     = {May},
  volume    = {249},
  number    = {5},
  pages     = {413--422}
}

@article{gronberg22_n,
  author    = {A. Groenberg and I. Henriksson and M. Stenman and A. G. Lindgren},
  title     = {Incidence of aphasia in ischemic stroke},
  journal   = {Neuroepidemiology},
  year      = {2022},
  month = {Mar.},
  volume    = {56},
  number    = {3},
  pages     = {174--182}
}

@article{palmer12_stroke,
  author    = {R. Palmer and P. Enderby and C. Cooper and N. Latimer and S. Julious and G. Paterson and M. Dimairo and S. Dixon and J. Mortley and R. Hilton and A. Delaney and H. Hughes},
  title     = {Computer therapy compared with usual care for people with long-standing aphasia poststroke: a pilot randomized controlled trial},
  journal   = {Stroke},
  year      = {2012},
month = {July},
  volume    = {43},
  number    = {7},
  pages     = {1904--1911}
}

@article{salis16_apha,
  author    = {C. Salis and F. Hwang},
  title     = {Digital technology and aphasia},
  journal   = {Aphasiology},
  year      = {2016},
  volume    = {30},
  number    = {2--3},
  pages     = {109--111},
}

@inproceedings{elizalde23_clap,
  author       = {B. Elizalde and S. Deshmukh and M. A. Ismail and H. Wang},
  title        = {CLAP: Learning Audio Concepts from Natural Language Supervision},
  booktitle    = {Proc. IEEE International Conference on Acoustics, Speech and Signal Processing},
  year         = {2023},
  month = {June},
  address      = {Rhodes Island, Greece},
}

@inproceedings{w2v2,
  title={wav2vec 2.0: A framework for self-supervised learning of speech representations},
  author={Baevski, Alexei and Zhou, Yuhao and Mohamed, Abdelrahman and Auli, Michael},
  booktitle={Proc. Annual Conference
                  on Neural Information Processing Systems},
  address={Virtual},
  pages={12449--12460},
  year={2020},
  month={Dec.},
}

@inproceedings{Sanh2019,
author = {Sanh, Victor and Debut, Lysandre and Chaumond, Julien and Wolf, Thomas},
booktitle = {Proc. 5th Workshop on Energy Efficient Machine Learning and Cognitive Computing @ NeurIPS 2019},
title = {{DistilBERT, a distilled version of BERT: smaller, faster, cheaper and lighter}},
address = {Vancouver, Canada},
year = {2019},
month = {Dec.}
}

@misc{ot25,
      title={A Differentiable Alignment Framework for Sequence-to-Sequence Modeling via Optimal Transport}, 
      author={Yacouba Kaloga and Shashi Kumar and Petr Motlicek and Ina Kodrasi},
      year={2025},
      eprint={2502.01588},
      archivePrefix={arXiv},
      primaryClass={cs.LG},
      url={https://arxiv.org/abs/2502.01588}, 
}

@misc{kingma2017adam,
      title={Adam: A Method for Stochastic Optimization}, 
      author={Diederik P. Kingma and Jimmy Ba},
      year={2017},
      eprint={1412.6980},
      archivePrefix={arXiv},
      primaryClass={cs.LG},
      url={https://arxiv.org/abs/1412.6980}, 
}

@misc{loshchilov2019,
      title={Decoupled Weight Decay Regularization}, 
      author={Ilya Loshchilov and Frank Hutter},
      year={2019},
      eprint={1711.05101},
      archivePrefix={arXiv},
      primaryClass={cs.LG},
      url={https://arxiv.org/abs/1711.05101}, 
}

@misc{wu2018un,
      title={Unsupervised Feature Learning via Non-Parametric Instance-level Discrimination}, 
      author={Zhirong Wu and Yuanjun Xiong and Stella Yu and Dahua Lin},
      year={2018},
      eprint={1805.01978},
      archivePrefix={arXiv},
      primaryClass={cs.CV},
      url={https://arxiv.org/abs/1805.01978}, 
}

@misc{liang2025,
      title={Selection of Layers from Self-supervised Learning Models for Predicting Mean-Opinion-Score of Speech}, 
      author={Xinyu Liang and Fredrik Cumlin and Victor Ungureanu and Chandan K. A. Reddy and Christian Schuldt and Saikat Chatterjee},
      year={2025},
      eprint={2508.08962},
      archivePrefix={arXiv},
      primaryClass={eess.AS},
      url={https://arxiv.org/abs/2508.08962}, 
}

@misc{baevski2020,
      title={wav2vec 2.0: A Framework for Self-Supervised Learning of Speech Representations}, 
      author={Alexei Baevski and Henry Zhou and Abdelrahman Mohamed and Michael Auli},
      year={2020},
      eprint={2006.11477},
      archivePrefix={arXiv},
      primaryClass={cs.CL},
      url={https://arxiv.org/abs/2006.11477}, 
}

@misc{parcollet2024,
      title={LeBenchmark 2.0: a Standardized, Replicable and Enhanced Framework for Self-supervised Representations of French Speech}, 
      author={Titouan Parcollet and Ha Nguyen and Solene Evain and Marcely Zanon Boito and Adrien Pupier and Salima Mdhaffar and Hang Le and Sina Alisamir and Natalia Tomashenko and Marco Dinarelli and Shucong Zhang and Alexandre Allauzen and Maximin Coavoux and Yannick Esteve and Mickael Rouvier and Jerome Goulian and Benjamin Lecouteux and Francois Portet and Solange Rossato and Fabien Ringeval and Didier Schwab and Laurent Besacier},
      year={2024},
      eprint={2309.05472},
      archivePrefix={arXiv},
      primaryClass={cs.CL},
      url={https://arxiv.org/abs/2309.05472}, 
}

@misc{babu2021xlsr,
      title={XLS-R: Self-supervised Cross-lingual Speech Representation Learning at Scale}, 
      author={Arun Babu and Changhan Wang and Andros Tjandra and Kushal Lakhotia and Qiantong Xu and Naman Goyal and Kritika Singh and Patrick von Platen and Yatharth Saraf and Juan Pino and Alexei Baevski and Alexis Conneau and Michael Auli},
      year={2021},
      eprint={2111.09296},
      archivePrefix={arXiv},
      primaryClass={cs.CL},
      url={https://arxiv.org/abs/2111.09296}, 
}

@article{Franco2025,
  author       = {J. Franco and B. Glize and M. Laganaro},
  title        = {Impact of immersive virtual reality compared to a digital static approach in word (re)learning in post-stroke aphasia and neurotypical adults: Lexical-semantic effects?},
  journal      = {Neuropsychologia},
  year         = {2025},
  month        = {Feb.},
  volume       = {208},
}

@article{Demierre2025,
  author    = {Cyrielle Demierre and Bertrand Glize and Marina Laganaro},
  title     = {Increase of phonological errors in dual-task conditions in patients with aphasia and neurotypical individuals: Impact of the verbal nature of the concurrent task},
  journal   = {Neuropsychologia},
  year      = {2025},
  month     = {May},
  volume    = {211},
}

@article{Macoir2016,
  author    = {Jean Macoir and Christine Gauthier and C{\'e}line Jean and Olivier Potvin},
  title     = {{BECLA, a new assessment battery for acquired deficits of language: Normative data from Quebec-French healthy younger and older adults}},
  journal   = {Journal of the Neurological Sciences},
  year      = {2016},
  month     = {Feb},
  volume    = {361},
  pages     = {220--228},
}

@misc{devlin2019,
      title={BERT: Pre-training of Deep Bidirectional Transformers for Language Understanding}, 
      author={Jacob Devlin and Ming-Wei Chang and Kenton Lee and Kristina Toutanova},
      year={2019},
      eprint={1810.04805},
      archivePrefix={arXiv},
      primaryClass={cs.CL},
      url={https://arxiv.org/abs/1810.04805}, 
}

@article{Mason_Nickels_2022,
  author       = {C. Mason and L. Nickels},
  title        = {Are single-word picture naming assessments a valid measure of word retrieval in connected speech?},
  journal      = {International Journal of Speech-Language Pathology},
  year         = {2022},
  volume       = {24},
  number       = {1},
  pages        = {97--109},
}

@article{Conroy_et_al_2018,
  author  = {P. Conroy and C. Sotiropoulou Drosopoulou and G. F. Humphreys and A. D. Halai and M. A. Lambon Ralph},
  title   = {{Time for a quick word? The striking benefits of training speed and accuracy of word retrieval in post-stroke aphasia}},
  journal = {Brain},
  year    = {2018},
  volume  = {141},
  number  = {6},
  pages   = {1815--1827}
}

\end{document}